\documentclass[journal]{IEEEtran}
\usepackage{graphicx}

\usepackage{amsmath}
\usepackage[table,xcdraw]{xcolor}
\usepackage{array}
\usepackage{caption}
\usepackage{floatrow}   
\usepackage{subcaption}
\usepackage{graphicx,xcolor} 
\usepackage{breqn}
\usepackage[framemethod=tikz]{mdframed}
\usepackage[tableposition=top]{caption}
\usepackage{float}
\floatstyle{plaintop}
\restylefloat{table}
\usepackage{arydshln}
\usepackage{hyperref}
\usepackage{color, colortbl}
\definecolor{Gray}{gray}{0.85}
\usepackage{booktabs}
\usepackage[normalem]{ulem}
\useunder{\uline}{\ul}{}

\usepackage{ifpdf}

\usepackage{cite}
\usepackage{cite}
\usepackage{amsmath,amssymb,amsfonts}
\usepackage{algorithmic}
\usepackage{graphicx}
\usepackage{graphicx}

\usepackage{amsmath}
\usepackage{array}
\usepackage{caption}
\usepackage{floatrow}   
\usepackage{subcaption}
\usepackage{graphicx,xcolor} 
\usepackage[colorinlistoftodos]{todonotes}

\usepackage{breqn}
\usepackage[framemethod=tikz]{mdframed}
\usepackage[tableposition=top]{caption}
\usepackage{float}
\floatstyle{plaintop}
\restylefloat{table}
\usepackage{arydshln}
\usepackage{hyperref}
\usepackage{pifont}
\usepackage{amssymb}
\usepackage{pifont}
\usepackage{ifpdf}
\usepackage{booktabs}
\usepackage{makecell}
\usepackage{sidecap}

\sidecaptionvpos{figure}{c}

\newcommand{\comment}[1]{}

\usepackage{ifpdf}

\usepackage{cite}

\begin{document}
	%
	\title{Bi-Directional Semi-Supervised Training\\ of Convolutional Neural Networks for \\Ultrasound Elastography Displacement Estimation}
	%
	%
	%

	\author{Ali~K. Z. Tehrani, Mostafa~Sharifzadeh, Emad Boctor and Hassan~Rivaz 
		
		\thanks{A. K. Z. Tehrani, Mostafa Sharifzadeh, and H. Rivaz are with the Department
			of Electrical and Computer Engineering, Concordia University, Canada,
			Emad Boctor is with the Department of Radiology and Radiological Science, John Hopkins University, United States,
			e-mail: A\_Kafaei@encs.concordia.ca,
			mostafa.sharifzadeh@concordia.ca,
			eboctor@jhmi.edu and 
			hrivaz@ece.concordia.ca
		}
		\thanks{}}
	
	%
	%

\markboth{IEEE TUFFC}%
{Shell \MakeLowercase{\textit{et al.}}: Bare Demo of IEEEtran.cls for IEEE Journals}
%



\maketitle

\begin{abstract}
	The performance of ultrasound elastography (USE) heavily depends on the accuracy of displacement estimation. Recently, Convolutional Neural Networks (CNN) have shown promising performance in optical flow estimation and have been adopted for USE displacement estimation. Networks trained on computer vision images are not optimized for USE displacement estimation since there is a large gap between the computer vision images and the high-frequency Radio Frequency (RF) ultrasound data. Many researchers tried to adopt the optical flow CNNs to USE by applying transfer learning to improve the performance of CNNs for USE. However, the ground truth displacement in real ultrasound data is unknown, and simulated data exhibits a domain shift compared to the real data and is also computationally expensive to generate. To resolve this issue, semi-supervised methods have been proposed wherein the networks pre-trained on computer vision images are fine-tuned using real ultrasound data. In this paper, we employ a semi-supervised method by exploiting the first and second-order derivatives of the displacement field for the regularization. We also modify the network structure to estimate both forward and backward displacements, and propose to use consistency between the forward and backward strains as an additional regularizer to further enhance the performance. We validate our method using several experimental phantom and \textit{in vivo} data. We also show that the network fine-tuned by our proposed method using experimental phantom data performs well on \textit{in vivo} data similar to the network fine-tuned on \textit{in vivo} data. Our results also show that the proposed method outperforms current deep learning methods and is comparable to computationally expensive optimization-based algorithms.  
\end{abstract}

\begin{IEEEkeywords}
	Ultrasound Elastography, Displacement Estimation, Convolutional Neural Network, Optical Flow, Semi-Supervised Training.
\end{IEEEkeywords}

%
\IEEEpeerreviewmaketitle

\section{Introduction}
Ultrasound (US) imaging has been increasingly used by researchers and clinicians in diagnosis and image guided intervention since it is less expensive and more portable compared to other imaging modalities. US Elastography (USE) is an imaging technique that detects viscoelastic properties of the tissue and has been found useful in different applications including ablation monitoring \cite{kling2018potential,lee2019evaluating} and breast lesion characterization \cite{hall2001vivo}. USE methods track the motion of the tissue and can be broadly categorized into two groups: dynamic and quasi-static elastography. In dynamic elastography, acoustic radiation force or an internal force is used to generate fast motions in the tissue \cite{bercoff2004supersonic}. In contrast, the motions in the tissue are slow in quasi-static elastography, and can be induced by simply pressing the probe by the operator (free-hand palpation) or using a robotic arm \cite{ophir1999elastography,varghese2000direct,schneider2012remote,rivaz2010real}. 

In free-hand palpation USE, the operator compresses the tissue by the probe usually in the axial direction. The US Radio Frequency (RF) data before and after the compression are compared to obtain the displacement map which indicates the movement of each individual sample. The displacement map is utilized to obtain the strain map which contains relative elasticity information. In this paper, we aim to estimate axial displacement for quasi-static free-hand palpation USE.      

Existing techniques for performing the displacement estimation step can be categorized into conventional and deep learning-based methods. Window-based \cite{varghese2000direct,nahiyan2015hybrid,luo2010fast,jiang2007parallelizable,mirzaei20203d} and optimization-based \cite{j2011recent,hashemi2017global,rivaz2010real,mirzaei2019combining} are two main groups of conventional methods that have been used widely for USE. Recent optimization-based methods have surpassed window-based ones \cite{hashemi2017global,mirzaei2019combining}. The main idea of the optimization-based methods is to determine an initial coarse estimation, usually by Dynamic Programming (DP) \cite{rivaz2010real}, followed by minimizing a regularized cost function to obtain the displacement map. OVERWIND is a recent algorithm that combines the window-based and optimization-based methods \cite{mirzaei2019combining}. OVERWIND employs L1 norm as the regularizer to preserve the sharpness of the displacements on the boundaries.

Deep learning-based methods, which have been recently proposed for USE, employ optical flow Convolutional Neural Networks (CNN) to obtain the displacement map. The first few works used the optical flow CNNs as black boxes for USE \cite{peng2020neural,peng2018convolution} or as the initial estimator for optimization-based methods instead of DP \cite{kibria2018gluenet,kibria2018global}. However, the computer vision images and US data are vastly different and the CNN architectures used for the former are not optimized for high-frequency RF data. Motivated to address this issue, we modified the well-known PWC-Net architecture  \cite{sun2018pwc} to be adapted to USE, considering the physics of RF data \cite{tehrani2020displacement}. We called the network Modified PWC-Net (MPWC-Net) and obtained substantially more accurate displacement compared to PWC-Net. In another work, we proposed MPWC-Net++ which was an improved version of MPWC-Net with a higher search range and more accurate output displacement \cite{tehrani2021mpwc}. These methods require a GPU to run efficiently, and can perform high frame-rate USE given the rapidly increasing computational power of GPUs. However, their main drawback is that they have a larger variance compared to conventional methods due to the fact that they are not regularized in contrast to conventional methods. Consequently, their strain images have lower overall quality compared to conventional methods \cite{tehrani2020displacement,tehrani2021mpwc,tehrani2020real}.                   

Unsupervised training was another avenue that has been followed by the researchers. Delaunay \textit{et al.} trained a U-Net using real US data \cite{delaunay2020unsupervised} and developed a recurrent network to deal with a sequence of frames \cite{delaunay2021unsupervised}. In \cite{tehrani2020semi}, we used a light network, referred to as LiteFlowNet \cite{hui2018liteflownet} and trained it in a semi-supervised fashion. We first used computer vision datasets with known ground truths to train the network using supervised techniques. In the next step, real US data was used to fine-tune the network using an unsupervised method. We substantially improved the strain image quality by using this technique without requiring a large amount of training data. 

\begin{figure*}[!t]
	\centering
	\includegraphics[width=0.72\textwidth]{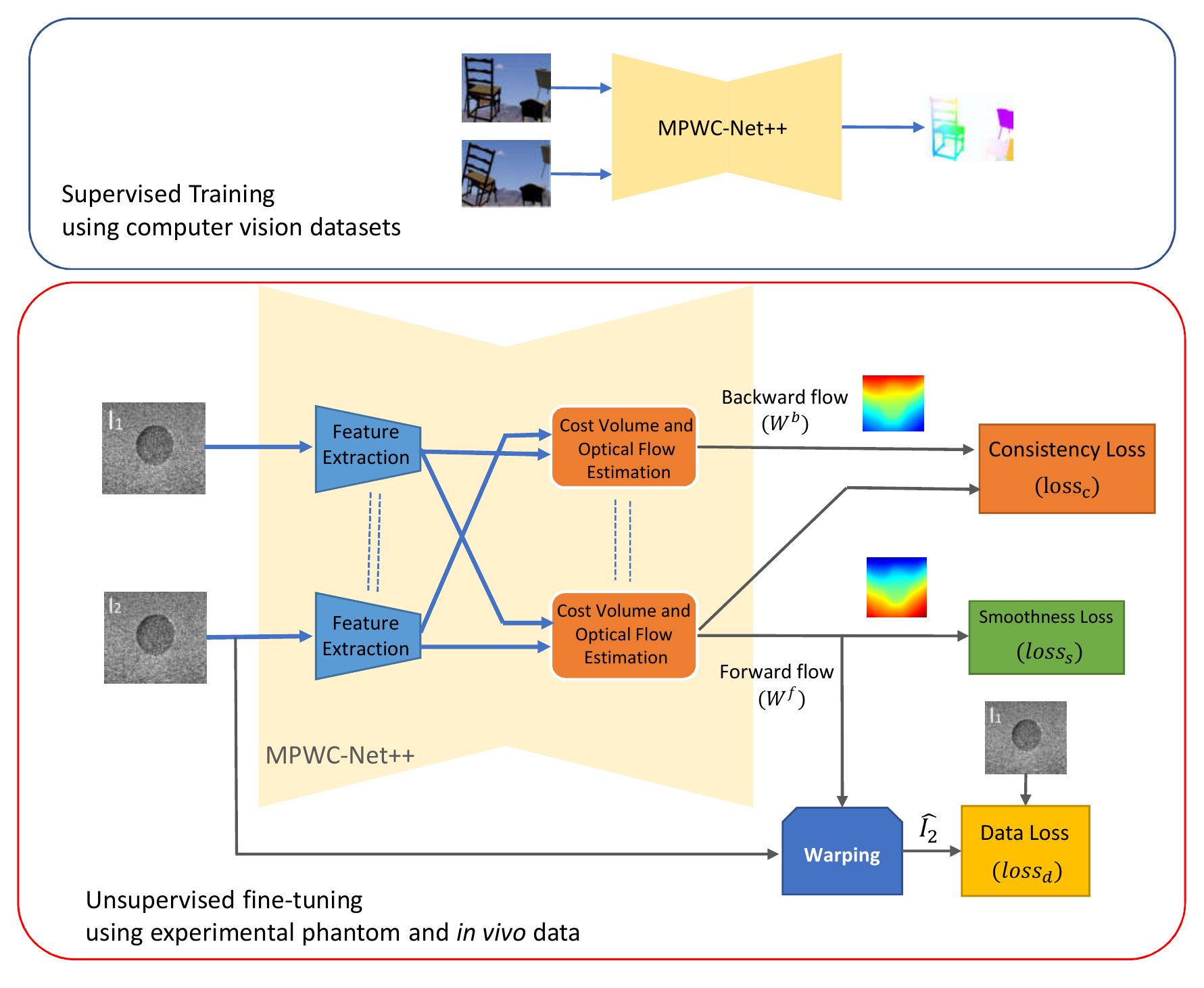}
	\centering
	\caption{Overview of the proposed semi-supervised training method. The network is first trained using computer vision datasets by supervised methods (the block on top). The network is fine-tuned by real US data using the proposed unsupervised training method (bottom block). The network structure is altered to be able to estimate both forward and backward flows. The layers connected by the dashed lines share weights. The cost volume and optical flow estimation blocks with shared weights are used to estimate both forward and backward flows.} 
	\label{fig:network}
\end{figure*} 

In this paper, we follow the semi-supervised training approach. The overview of the method is depicted in Fig. \ref{fig:network}. We first employ computer vision datasets to train the network in a supervised fashion. We use MPWC-Net++, which has shown high performance in USE. We then fine-tune the network by real US data, and extend our idea of semi-supervised method by proposing bi-directional unsupervised fine-tuning. We change the structure of MPWC-Net++ to estimate both forward and backward displacements, which is more efficient than running the network two times to estimate them. Consistency loss is also proposed which is obtained by comparing forward and backward strains. Furthermore, we shed light on the choice of weights for regularization by relating some of them to others. We demonstrate the high-performance of the proposed methods using experimental phantom and \textit{in vivo} data.
%

\section{Material and Method}

\subsection{Deep Supervised CNNs}
In this section, several CNNs used in USE displacement estimation are explained. 
\subsubsection{PWC-Net}
Sun \textit{et al.} proposed using cost volume and warping of the features for optical flow estimation \cite{sun2018pwc}. They achieved state-of-the-art performance in different computer vision datasets. PWC-Net used a pyramidal structure, in which the optical flow was estimated in different resolution levels. At each pyramid level, the features of the second image were warped by the estimated flow of the previous pyramid to reduce the flow in the next pyramid level. In the next step, cost volume was employed to compare features of the fixed and warped moved images. In the last step, the optical flow was estimated and used in the next pyramid. Using pyramid structure resulted in the reduction of the number of learnable weights and improved the performance of optical flow estimation \cite{sun2018pwc}. Recently, a variant of PWC-Net called Iterative Residual Refinement PWC-Net (IRR-PWC-Net) was proposed \cite{hur2019iterative}. This network reduced the number of learnable weights even further by using the optical flow estimation block iteratively.

\subsubsection{MPWC-Net}
PWC-Net was proposed for computer vision images originally, and was not well suited to the high-frequency RF data. We modified the structure of PWC-Net and proposed MPWC-Net for USE \cite{tehrani2020displacement}. We removed the strides of the first feature extraction layer to preserve high-frequency information in RF data. In order to avoid failure of the network in low pyramid levels, where RF data does not have enough information due to downsampling, we used envelope and B-mode images as additional input channels. We obtained competitive performance with conventional optimization-based methods. The code and the simulation dataset for fine-tuning are available online at \href{http://code.sonography.ai}{code.sonography.ai}.
\subsubsection{MPWC-Net++}
MPWC-Net had a low displacement range since strides of the first feature extraction layer were removed. In addition, we showed that the real displacement range is much lower than the theoretical one since only a small quantity of training data have high displacement ranges. Therefore, the network is not trained enough to deal well with large displacements, and the predicted flows are noisy in this condition. To address these problems, instead of removing both strides, we only removed one of them and kept the other one. Also, the search range of the cost volume was increased from 4 to 5 in each pyramid level. By doing these modifications, the network had higher search range and could work better for large displacements. Furthermore, we used IRR-PWC-Net \cite{hur2019iterative} since it has a more efficient structure. Although we applied these modifications to improve the performance in the USE application, they also led to an improvement even for the computer vision dataset. \cite{tehrani2021mpwc}.

However, the improvements came with costs. Due to modifying the network structure, we had to train the network from scratch (MPWC-Net did not require training from scratch). The training itself was slower compared to the original PWC-Net for two reasons: first, having larger feature maps after removing strides; second, increasing the cost volume search range. The MPWC-Net++ was trained from scratch using computer vision images while LiteFlowNet used in \cite{tehrani2020semi} was pre-trained and we only fine-tuned it using real US data. Training from scratch takes 840 hours using an NVIDIA Tesla P6 GPU, substantially more time compared to the fine-tuning since the network is randomly initialized and a large dataset (for MPWC-Net++ 22000 pairs) is employed to train the network. The network weights are publicly available online at \href{http://code.sonography.ai}{code.sonography.ai}.

\subsection{Semi-supervised Method}
Simulation data does not model non-linear or multiple scattering effects present in real US data \cite{delaunay2020unsupervised,tehrani2020semi}. Therefore, we proposed to use real US data for fine-tuning. This method was semi-supervised since we first used a pre-trained network trained on computer vision images by supervised methods. In the next step, real US data were used to fine-tune the network.

The moved image was warped with the forward flow and compared with the first image which is called photometric loss. This loss alone resulted in noisy displacements; therefore, inspired by the physics of RF data, we proposed using the first second-order derivatives of displacement in the axial and lateral direction as the regularization. The detailed description of the loss function can be found in \cite{tehrani2020semi}.

In order to preserve the information of high-frequency RF data, we were not allowed to downsample images and had to use large image sizes during unsupervised fine-tuning which is challenging due to GPU memory limitations. We used a light network (LiteFlowNet \cite{hui2018liteflownet}) and gradient checkpointing \cite{hung2019optimization} to be able to train the network on our GPU (Nvidia TITAN V with 12 GB of RAM). Furthermore, we limited the training to only forward flow and the backward flow was used to detect occluded regions. 

\subsection{Proposed Method}
Let $I_1,I_2\in \mathbb{R}^{3\times W \times H}$ denote the fixed and moved images having 3 channels with width $W$ and height $H$, $W^f\in \mathbb{R}^{2\times W \times H}$ and $W^b\in \mathbb{R}^{2\times W \times H}$ denote forward flow ($I_1\rightarrow I_2$) and backward flow ($I_2\rightarrow I_1$), respectively. The data loss function for unsupervised fine-tuning can be defined as \cite{tehrani2020semi}:
\begin{equation}
	\label{eq:loss_d}
	loss_d = \Phi (I_1-\hat{I_2})_{w\times w} 
\end{equation}
where $\hat{I_2}$ is the second image warped by the $W^f$ and unlike \cite{tehrani2020semi}, a window of size $(w\times w)$ around the sample of interest is selected to compute the loss to reduce the noise caused by interpolation step of warping operation (here we use a $3\times3$ window). $\Phi$ is the Charbonnier loss that has been widely used in unsupervised optical flow training \cite{jonschkowski2020matters} and defined in Eq \ref{Charbonnier} \cite{barron2019general}.
\begin{equation}
	\label{Charbonnier}
	\Phi(x)=|(x^2+\varepsilon)^\alpha|_1    
\end{equation} 
where $|.|_1$ denotes the L1 norm and $\alpha$ can be altered to give different importance to $x$. We used $\alpha=0.5$ for the data loss (would be L1 norm) and $\alpha=0.2$ for smoothness and consistency loss to emphasize small values of $x$. It should be mentioned that inspired by \cite{tehrani2020real}, RF data, the envelope and imaginary part of Hilbert transform of RF data are utilized as 3 separate channels of input images. 

In order to have a smooth displacement field, the first-order derivatives of the displacements in axial and lateral directions are used for regularization:
\begin{equation}
	\label{eq:loss_s1}
	\begin{gathered}
		loss_s^1 = \lambda _{11} \Phi\ (\frac{\partial W^f_a}{\partial a}  -<\frac{\partial }{\partial a}W^f_a>)+ \lambda _{12} \Phi\ (\frac{\partial W^f_a}{\partial l}) + \\ \lambda _{21} \Phi\ (\frac{\partial W^f_l}{\partial a} )+ \lambda _{22} \Phi\ (\frac{\partial W^f_l}{\partial l}) 
	\end{gathered}  
\end{equation}    
where $W^f_a$, $W^f_l$, $\frac{1}{\partial a}$, $\frac{1}{\partial l}$ and $\lambda$ denote axial, lateral displacements, the derivative in axial and lateral directions and their corresponding weights, respectively. The axial derivative of the axial displacement is subtracted by its mean ($<.>$ denotes the mean value) to reduce the bias of the regularization similar to \cite{mirzaei2019combining}.

The second-order derivatives of the displacements have been found useful for USE \cite{tehrani2020semi,delaunay2020unsupervised,ashikuzzaman2021combining}. Hence, they can be used to regularize the displacements:

\begin{equation}
	\label{eq:loss_s2}
	\begin{gathered}
		loss_s^2 = \lambda _{31} \Phi\ (\frac{\partial^2 W^f_a}{\partial^2 a}) +  \lambda _{32} \Phi\ (\frac{\partial^2 W^f_a}{\partial a \partial l}) + \\
		\lambda _{41} \Phi\ (\frac{\partial^2 W^f_l}{\partial l \partial a})+\lambda _{42} \Phi\ (\frac{\partial^2 W^f_l}{\partial^2 l}) 
	\end{gathered}  
\end{equation}
Unlike \cite{delaunay2021unsupervised} that used the first and second-order derivatives of only axial displacement, we used the first and second-order derivatives of both axial and lateral displacement in both directions. The second-order derivatives do not introduce bias but they require higher weights to be as effective as the first-order derivatives.
\subsubsection{Hyper-Parameter Tuning}    
It can be seen that there are eight hyper parameters that we need to set before training the network. In our recent work \cite{tehrani2020semi}, we set them empirically, while in this paper, we tried to reduce the number of hyper-parameters by relating some of the them to others using US principles. 

	
	The distance between two adjacent samples in the axial and lateral direction is also vastly different since the sampling frequencies and the number of samples is widely dissimilar. The distance between two adjacent samples in the axial direction can be obtained by $c/(2f_s)$, where $c$ denotes sound speed and $f_s$ is the sampling frequency. A rough approximation of the lateral distance between two samples would be the distance between two adjacent A-lines which is much larger than the axial distance in a typical US image. Therefore, $\lambda _{12}$ and $\lambda _{22}$ must be several times smaller than $\lambda _{11}$ and $\lambda _{21}$.  
	
	As noted in \cite{ashikuzzaman2021combining}, the second-order derivatives are much smaller than the first-order ones. Therefore, to be as effective as the first-order derivatives, their weight should be several times larger than the first-order derivatives. We set this weight to be $\lambda_{31}=5\lambda_{11}$. We also set the lateral derivative weights similar to the first-order derivatives ($\lambda_{41}=\beta\lambda_{31}$). Finally, the smoothness regularizer can be written as:
	
	\begin{equation}
		\label{eq:loss_s}
		\begin{gathered}         
			loss_s =  \Phi\ (\frac{\partial W^f_a}{\partial a}  -<\frac{\partial }{\partial a}W^f_a>)  
			+ \beta \Phi\ (\frac{\partial W^f_a}{\partial l})+\\ 0.5\Phi\ (\frac{\partial W^f_l}{\partial a} )  +  0.5\beta\Phi\ (\frac{\partial W^f_l}{\partial l})+ \\ 5\biggl \{\Phi\ (\frac{\partial^2 W^f_a}{\partial^2 a}) + \beta \Phi\ (\frac{\partial^2 W^f_a}{\partial a\partial l})\biggr. +\\
			\left. 0.5\Phi\ (\frac{\partial^2 W^f_l}{\partial{l}\partial{a}})+0.5\beta\Phi\ (\frac{\partial^2 W^f_l}{\partial^2 l}) \right \}
		\end{gathered}  
	\end{equation}
	where $\beta$ depends on the ratio of the sampling frequency in axial and lateral directions. Setting the weights does not require an exact calculation of the sampling frequencies and should be less than 1; we set this hyper parameter to 0.1. Finally, for incompressible materials, the Poisson’s ratio is approximately 0.5 \cite{chen1996young}, which means that the strain in the lateral direction is half of the axial one. Therefore, $\lambda _{21}$ can be substituted by $0.5\lambda _{11}$. 
	It should be noted that the explained method to tune the weights of the regularizers is only a rough estimate of the optimal values; therefore, the training is not too sensitive to the variations of these weights and even changing the weights by as much as 100\% yields similar results.  
	
	\subsubsection{Bi-directional Strain Consistency}
	In this paper, inspired by recent unsupervised methods in optical flow estimation \cite{meister2018unflow}, we proposed to utilize forward and backward consistency in addition to the data and smoothness losses. In unsupervised optical flow methods, the difference between forward and backward displacement was used for consistency loss \cite{meister2018unflow}. 
	
	Strain images are often showed in USE as a surrogate of the elastic modulus. Therefore, it would be useful to utilize the derivatives of the displacements for the consistency loss. Assuming a uniform tissue, the estimated forward and backward axial strain can be written as:
	\begin{equation}
		\label{eq:cons0}
		\begin{gathered}         
			\varepsilon^f = \varepsilon_{gt} + \mathcal{N}(\mu,\,\sigma^{2})\\
			\varepsilon^b = -\varepsilon_{gt} + \mathcal{N}(-\mu,\,\sigma^{2})
		\end{gathered}
	\end{equation}
	where we assumed that the strain true value is $\varepsilon_{gt}$ and the error is modeled by a normal distribution with the bias and variance of $\mu$ and $\sigma^{2}$, respectively. The forward and backward consistency loss can be defined as:
		
		\begin{equation}
			\label{eq:cons2} 
			loss_c = \Phi(\varepsilon^f+\varepsilon^b)  
		\end{equation}
		Substituting Eq \ref{eq:cons0} into Eq \ref{eq:cons2}, yields:
		\begin{equation}
			\label{eq:cons3} 
			loss_c = \Phi(\mathcal{N}(0,\,\sigma^{2}/ 2))
		\end{equation}
		This equation indicates that minimizing the $loss_c$, results in reducing the variance in estimation of forward and backward strains. Similar to the smoothness loss ($loss_s$), the strain in both axial and lateral directions were employed which can be written as:
		\begin{equation}
			\label{eq:cons4} 
			loss_c = \Phi\ (\frac{\partial W^f_a}{\partial a}+\frac{\partial W^b_a}{\partial a})+ 0.5\beta \Phi\ (\frac{\partial W^f_l}{\partial l}+\frac{\partial W^b_l}{\partial l})  		
		\end{equation}
		where we used the same weights of the smoothness loss.
		
		By using the loss functions defined in Eq \ref{eq:loss_d},\ref{eq:loss_s}, and \ref{eq:cons4}, the total loss function can be written as:
		\begin{equation} 
			loss = loss_d + \lambda loss_s + \gamma loss_c
		\end{equation}
		
		Thanks to reducing the number of hyper-parameters, only $\lambda$ and $\gamma$ should be tuned for the training which can be done based on the training data. Too large values of $\lambda$ and $\gamma$ lead to a blurry strain image, while too small values result in noisy strain images. We set these hyper-parameters ($\lambda=0.03$ and $\gamma=0.05$) by visually inspecting the strain images of the validation set after training with different values of $\lambda$ and $\gamma$. It is worth mentioning that similar data and smoothness loss can be used for the backward flow. However, the consistency loss that we added has a similar behavior since it tries to make the backward strain close to the inverse of the forward one. 
		
		Adding the consistency loss demands high memory since both forward and backward flows are used for backpropagation. We utilized an NVIDIA A100 GPU with 40 GB of RAM to be able to train using the proposed loss function and simultaneously avoid downsampling to preserve RF data high frequency information. Since training this network might be infeasible for some researchers, we have made the networks weights available online at \href{http://code.sonography.ai}{code.sonography.ai}.

		\subsection{Data Collection}
		\subsubsection{Experimental Phantom}
		We used a tissue-mimicking breast phantom made from Zerdine (Model 059, CIRS: Tissue Simulation \& Phantom Technology, Norfolk, VA) for data collection. The phantom contained a number of hard inclusions. The background has elastic modulus of 20 kPa and the inclusions have at least twice elastic modulus of the background. This phantom was utilized to obtain training and test data. We made sure that different parts of the phantom were imaged for training and testing to avoid data leakage. 
		
		We employed Alpinion E-Cube R12 research US machine (Bothell,
		WA, USA) for training and test. The L3-12H linear array probe with the center frequency of 10 MHz and the sampling frequency of 40 MHz  was utilized for image acquisition.

		\subsubsection{\textit{In vivo} data}      
		\textit{In vivo} data was collected at Johns Hopkins Hospital using a research Antares Siemens system by a VF 10-5 linear array. The sampling frequency was 40 MHz and the center frequency was 6.67 MHz. Data was collected from patients with liver cancer during open-surgical RF thermal ablation. For more information, please see \cite{rivaz2010real}. The institutional review board approved the study with the consent of the patients.
		\subsection{Training schedule}
		We first collected data using the Alpinion machine from the breast phantom. We then selected image pairs having maximum axial displacement larger than 3 pixels. In total, 2200 image pairs were used for training. The networks were trained for 40 epochs and the learning rate was set to 30e-6 which reduced by 1/2 every 10 epochs. For \textit{in vivo} data results, we also fine-tuned the network trained by the bi-directional method using \textit{in vivo} data. This network was fine-tuned using 500 \textit{in vivo} image pairs for 20 epochs and the learning rate was 20e-6 and reduced by a factor of 2 every 5 epochs. In our experiments, we named this network as Bi-directional Unsupervised+ft.  
		\subsection{Compared methods}
		The compared methods are listed below.
		
		1) OVERWIND is an optimization-based method that estimates sub-pixel displacement. This method requires the initial displacement which is obtained by DP\cite{rivaz2010real}. OVERWIND considers a window around each sample and uses total variation for the regularization. This method obtains high quality strain images \cite{mirzaei2019combining}. 
		
		2) The recently proposed network, MPWC-Net++ which is the modified version of MPWC-Net. This network is only trained on computer vision images and no training on US data is done \cite{tehrani2021mpwc}.

		3) We fine-tune MPWC-Net++ using the unsupervised technique without the consistency loss. Also similar to \cite{tehrani2020semi}, it has the second-order derivatives only in the direction of the displacement ($\lambda _{32}, \lambda _{41} = 0$). In this case, the unsupervised fine-tuning would be similar to the semi-supervised method \cite{tehrani2020semi} with some minor improvements. The difference is that a better network (MPWC-Net++) with more suitable regularization weights is employed.
		
		4) Our proposed bi-directional unsupervised fine-tuning method. In this method, the consistency loss is added to the unsupervised loss function and all the second-order derivatives are employed in the smoothness loss. 
		
		5) For \textit{in vivo} data section, we also fine-tune the bi-directional network using \textit{in vivo} data. 
		
		We compare our bi-directional semi-supervised method with recent methods in USE: OVERWIND is a high-performance and non-deep learning method. MPWC-Net++ is one of the best networks used for USE without training on US data. The unsupervised variant of MPWC-Net++ combines this high-performance network with the unsupervised fine-tuning \cite{tehrani2020semi}.

		\section{Results}

		\subsection{Quantitative metrics}
		Contrast to Noise Ratio (CNR) and Strain Ratio (SR) are two popular metrics that have been used to assess the performance of the elastography methods. These metrics are suitable for experimental phantom and \textit{in vivo} data where the ground truth strain is unknown. Windows around the target and background are chosen to compute these metrics. CNR and SR can be obtained by \cite{ophir1999elastography}: 
		\begin{equation}
			\label{Eq:SRCNR}
			SR =\frac{\overline{s}_{t}}{\overline{s}_{b}},\quad \quad CNR = \sqrt{\frac{2(\overline{s}_{b}-\overline{s}_{t})^{2}}{{\sigma _{b}}^{2}+{\sigma _{t}}^{2}}},
		\end{equation}	
		where $\overline{s}_{t}$ and $\overline{s}_{b}$ are the mean of the strain in the selected target and background regions, respectively. The $\sigma _{t}$ and $\sigma _{b}$ are the standard deviations of the target and the background regions. Assuming that the target strain to be lower than the background region, lower SR represents a higher difference between the mean strain value of the target and the background region, and is therefore desired. CNR provides some notion of detectability, which combines the difference in mean between two samples is detectable and their background noise.

		Instead of selecting the target and background windows and calculating a single value for SR and CNR, thousands of SR and CNR values are calculated as follows. First, two large background and target windows are selected. We then select small patches within these windows, and calculate SR and CNR for different combinations of target and background windows. The number of these small patches should be large enough to produce statistically reliable estimates of SR and CNR. It should be mentioned that we selected large windows in regions with uniform strains (as marked in the figures) and small overlapping patches inside these windows to compute the CNR and SR. We then reported the mean and standard deviation of these values similar to our previous works \cite{tehrani2020displacement,tehrani2020semi,tehrani2021mpwc}.

		\subsection{Experimental Phantom Results}
\begin{figure}[!t]
				\centering
				\includegraphics[width=0.95\textwidth]{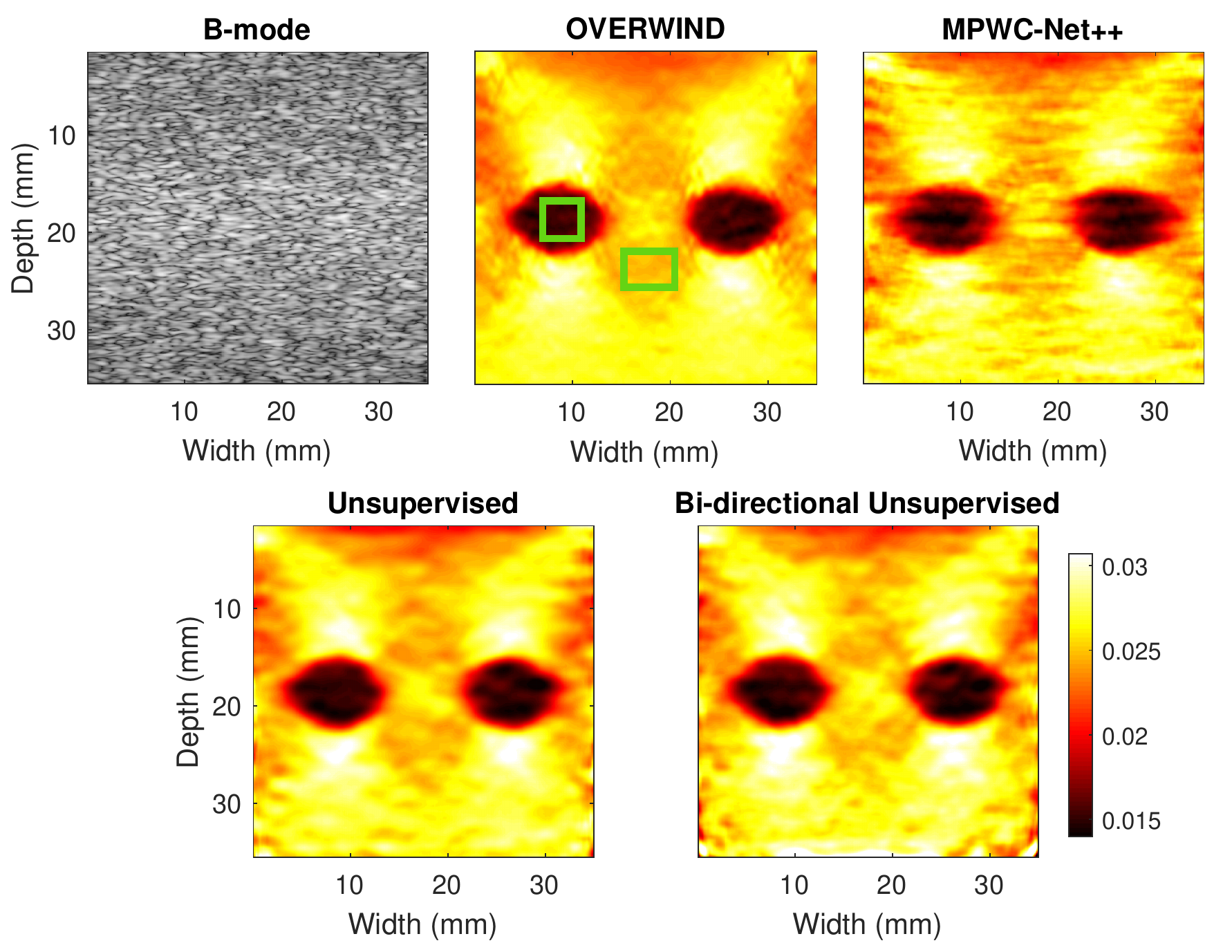}
				\centering
				\caption{Phantom Result 1 with the maximum strain value of 3$\%$. Green boxes indicate windows for computing CNR and SR.}
				\label{fig:phantom1}
			\end{figure}

\begin{figure}[!t]
				\centering
				\includegraphics[width=0.95\textwidth]{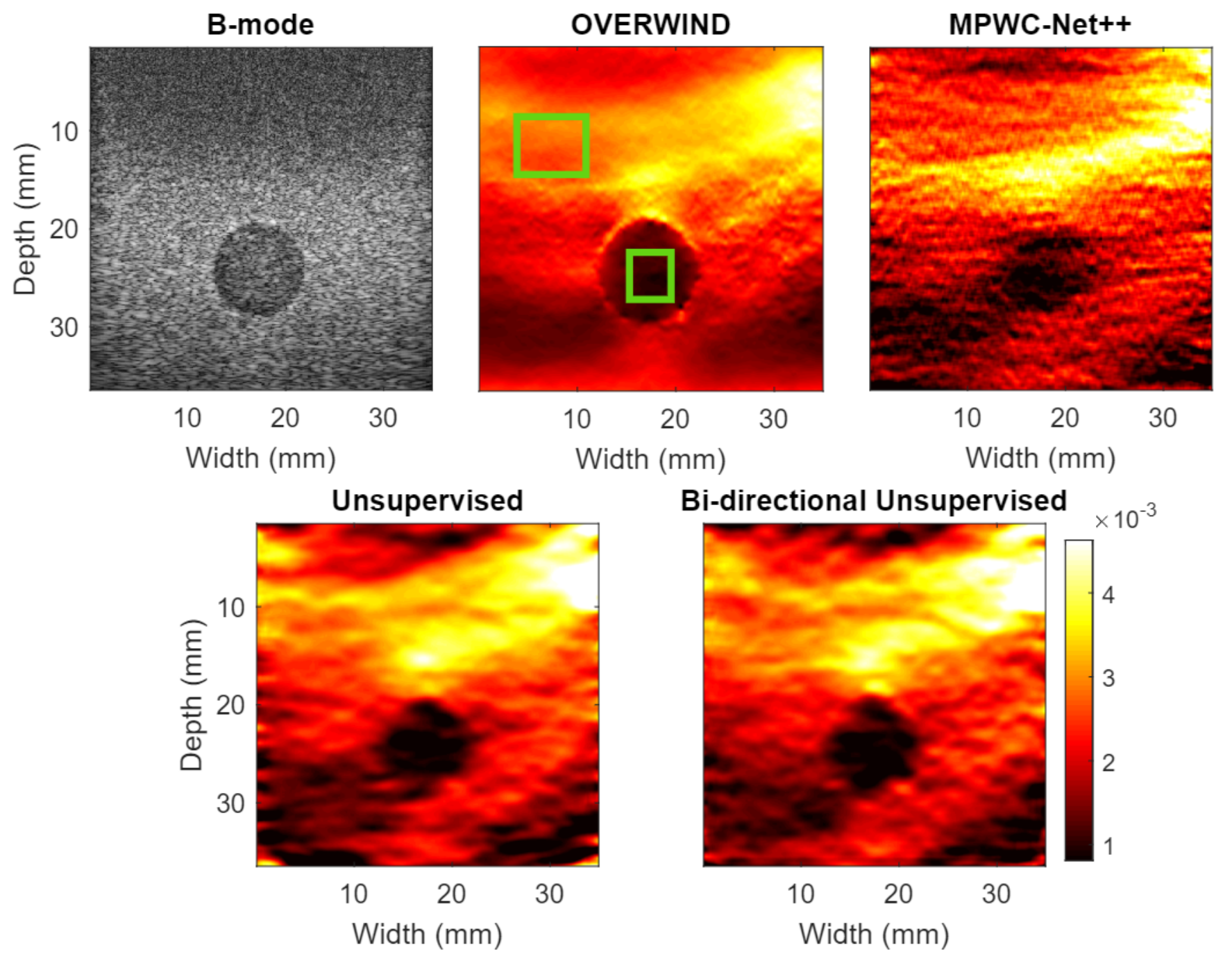}
				\centering
				\caption{Phantom Result 2 with the maximum strain value of 0.4$\%$. Green boxes indicate windows for computing CNR and SR.}
				\label{fig:phantom2}
			\end{figure}

\begin{figure}[!t]
				\centering
				\includegraphics[width=0.95\textwidth]{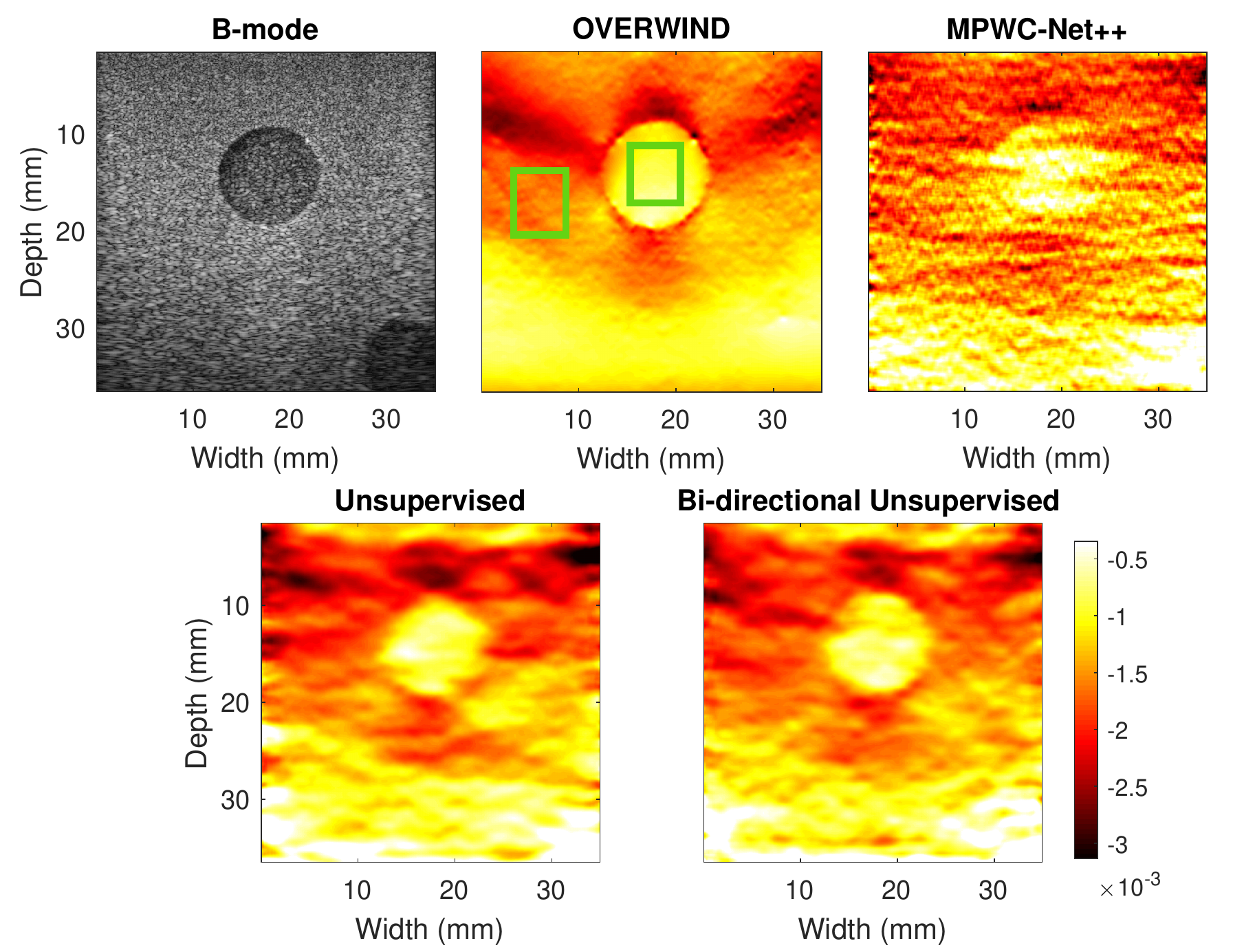}
				\centering
				\caption{Phantom Result 3 with the maximum strain value of 0.3$\%$. Green boxes indicate windows for computing CNR and SR.}
				\label{fig:phantom3}
			\end{figure}

\begin{figure}[!t]
				\centering
				\includegraphics[width=0.95\textwidth]{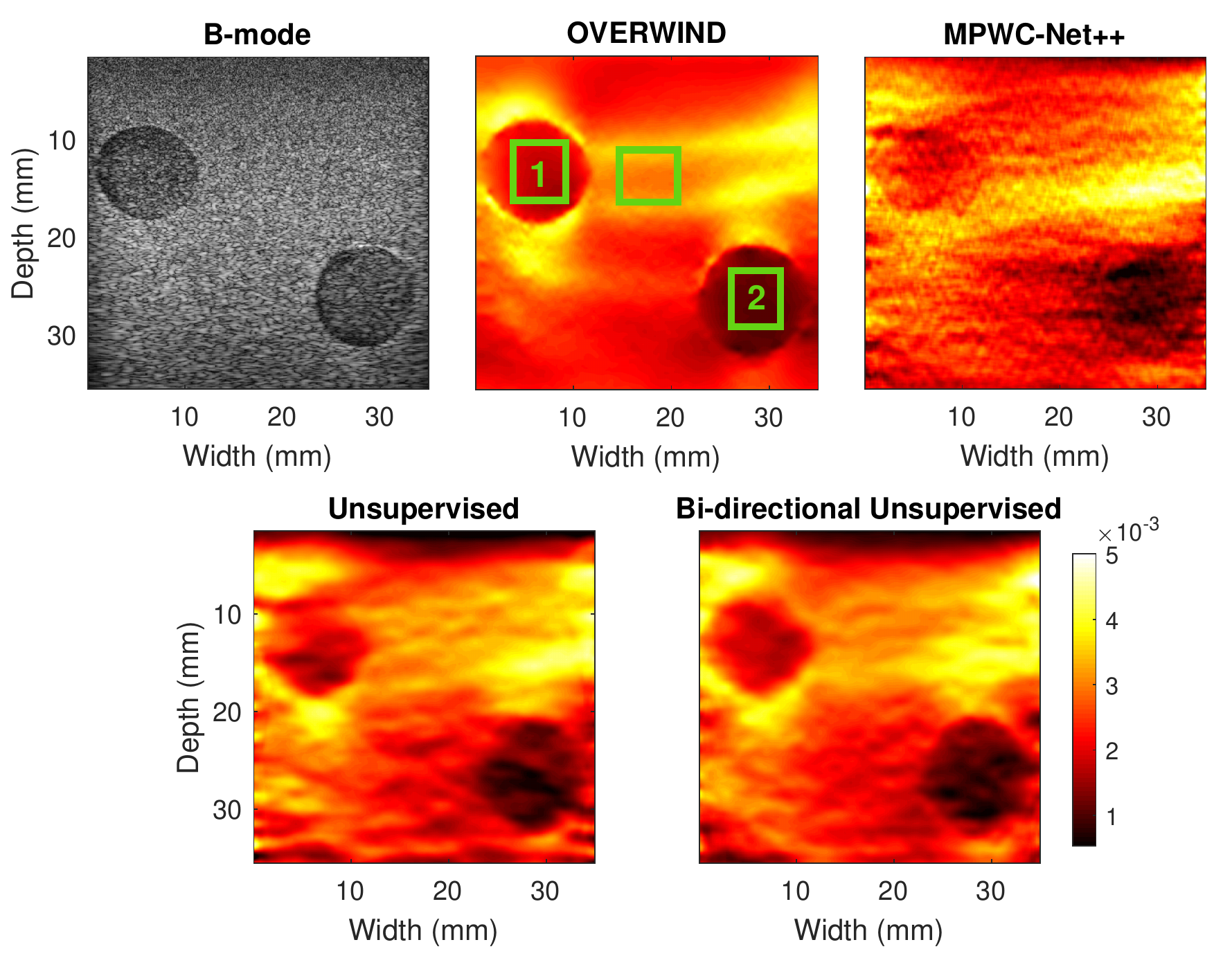}
				\centering
				\caption{Phantom Result 4 with two inclusions having different elasticity with the maximum strain value of 0.5$\%$. Green boxes indicate windows for computing CNR and SR.}
				\label{fig:phantom4}
			\end{figure}

		The results of different parts and compression levels of the experimental phantom are shown in Figs. \ref{fig:phantom1},\ref{fig:phantom2},\ref{fig:phantom3} and \ref{fig:phantom4}. In Fig. \ref{fig:phantom1}, the inclusions are not visible in the B-mode images, while they can be detected by USE methods. Comparing deep learning methods, unsupervised training substantially improves the strain quality of MPWC-Net++. Our proposed bi-directional method obtains similar or higher quality strain images compared to the unsupervised method, and substantially better results in all experiments compared to MPWC-Net++, especially when the compression is low (for example, Fig. \ref{fig:phantom3} and  \ref{fig:phantom4}). It can be seen that for those images MPWC-Net++ does not provide a clear image of the inclusion, while the proposed method obtains the highest quality strain images among the compared deep learning methods. OVERWIND obtains high-quality strain images and the proposed method performs comparably to OVERWIND in terms of the strain quality.

		\begin{table*}[]
			\caption{CNR results (higher is better). The bold font highlights the best, and the underline indicates the best deep learning results. Numbers marked with asterisks indicate results that are not statistically significant (\textit{p}-value$>0.01$), e.g. OVERWIND and the proposed method in Fig. \ref{fig:phantom1}.}
			\label{tab:phantom_CNR}
			\resizebox{0.8\textwidth}{!}{
				\begin{tabular}{@{}cclclclclcl@{}}
					\toprule
					&
					\multicolumn{2}{c}{Fig. \ref{fig:phantom1}} &
					\multicolumn{2}{c}{Fig. \ref{fig:phantom2}} &
					\multicolumn{2}{c}{Fig. \ref{fig:phantom3}} &
					\multicolumn{2}{c}{Fig. \ref{fig:phantom4} (1)} &
					\multicolumn{2}{c}{Fig. \ref{fig:phantom4} (2)} \\ \midrule
					OVERWIND &
					\multicolumn{2}{c}{27.26$\pm$4.27$^\ast$} &
					\multicolumn{2}{c}{\textbf{25.28$\pm$6.62}} &
					\multicolumn{2}{c}{12.40$\pm$2.38} &
					\multicolumn{2}{c}{\textbf{12.83$\pm$5.35$^\ast$}} &
					\multicolumn{2}{c}{\textbf{27.84$\pm$8.82}} \\ \midrule \addlinespace[0.19cm]
					MPWC-Net++ &
					\multicolumn{2}{c}{12.02$\pm$1.59} &
					\multicolumn{2}{c}{8.74$\pm$1.85} &
					\multicolumn{2}{c}{5.12$\pm$1.11} &
					\multicolumn{2}{c}{4.83$\pm$2.09} &
					\multicolumn{2}{c}{7.82$\pm$3.66} \\ \addlinespace[0.17cm]
					Unsupervised &
					\multicolumn{2}{c}{{\ul \textbf{31.78$\pm$7.47}}} &
					\multicolumn{2}{c}{15.12$\pm$4.12} &
					\multicolumn{2}{c}{9.94$\pm$1.64} &
					\multicolumn{2}{c}{8.67$\pm$4.09} &
					\multicolumn{2}{c}{14.37$\pm$4.27} \\ \addlinespace[0.17cm]
					Bi-directional Unsupervised &
					\multicolumn{2}{c}{27.71$\pm$5.20$^\ast$} &
					\multicolumn{2}{c}{{\ul 17.19$\pm$4.45}} &
					\multicolumn{2}{c}{{\ul \textbf{13.84$\pm$4.49}}} &
					\multicolumn{2}{c}{{\ul 12.82$\pm$4.83$^\ast$}} &
					\multicolumn{2}{c}{{\ul 21.40$\pm$3.69}} \\ \bottomrule \addlinespace[0.17cm]
			\end{tabular}}
		\end{table*}

			\begin{table*}[]
				\caption{SR($\%$) results (lower is better). The bold font highlights the best, and the underline indicates the best deep learning results. Numbers marked with asterisks  and star indicate results that are not statistically significant (\textit{p}-value$>0.01$).}
				\label{tab:phantom_SR}
				\resizebox{0.8\textwidth}{!}{
					\begin{tabular}{@{}cclclclclclcl@{}}
						\toprule
						&
						\multicolumn{2}{c}{Fig. \ref{fig:phantom1}} &
						\multicolumn{2}{c}{Fig. \ref{fig:phantom2}} &
						\multicolumn{2}{c}{Fig. \ref{fig:phantom3}} &
						\multicolumn{2}{c}{Fig. \ref{fig:phantom4} (1)} &
						\multicolumn{2}{c}{Fig. \ref{fig:phantom4} (2)} \\ \midrule
						OVERWIND &
						\multicolumn{2}{c}{62.26$\pm$0.71} &
						\multicolumn{2}{c}{40.21$\pm$3.2} &
						\multicolumn{2}{c}{50.04$\pm$3.08$^\star$} &
						\multicolumn{2}{c}{61.78$\pm$5.74$^\ast$} &
						\multicolumn{2}{c}{36.56$\pm$2.00} \\ \midrule \addlinespace[0.19cm]
						MPWC-Net++ &
						\multicolumn{2}{c}{61.56$\pm$1.70$^\ast$} &
						\multicolumn{2}{c}{35.08$\pm$6.26} &
						\multicolumn{2}{c}{49.29$\pm$4.41$^\ast$} &
						\multicolumn{2}{c}{69.43$\pm$8.46} &
						\multicolumn{2}{c}{40.20$\pm$10.3}  \\ \addlinespace[0.17cm]
						Unsupervised &
						\multicolumn{2}{c}{61.27$\pm$0.67$^\ast$} &
						\multicolumn{2}{c}{{\ul \textbf{25.89$\pm$3.97}}} &
						\multicolumn{2}{c}{48.15$\pm$6.16$^\star$$^\ast$} &
						\multicolumn{2}{c}{60.96$\pm$9.67$^\ast$} &
						\multicolumn{2}{c}{32.06$\pm$7.47}  \\ \addlinespace[0.17cm]
						Bi-directional Unsupervised &
						\multicolumn{2}{c}{{\ul \textbf{59.24$\pm$0.50}}} &
						\multicolumn{2}{c}{28.12$\pm$4.12} &
						\multicolumn{2}{c}{{\ul \textbf{45.82$\pm$2.53}}} &
						\multicolumn{2}{c}{{\ul \textbf{58.79$\pm$5.72}}} &
						\multicolumn{2}{c}{{\ul \textbf{30.32$\pm$2.90}}} \\ \bottomrule
					\end{tabular}
				}
			\end{table*}
			
			The quantitative results are listed in Table \ref{tab:phantom_CNR} (for CNR) and \ref{tab:phantom_SR} (for SR). In terms of CNR, our proposed method substantially increases the CNR of MPWC-Net++ and outperforms the unsupervised method in most cases. To be more specific, the bi-directional unsupervised method increases the CNR of MPWC-Net++ from 12.02, 8.74, 5.12, 3.73 and 10.25 to 27.71, 17.19, 13.84, 12.82, and 21.40, respectively. It also obtains CNR values close to OVERWIND or even better (in Fig. \ref{fig:phantom3}). 
			
			The SR results agree with \cite{tehrani2020displacement} where we showed that MPWC-Net has better SR compared to another optimization-based method. In most cases (except Fig. \ref{fig:phantom2})), bi-directional unsupervised method has the better SR and close to the lowest ones in that case. 
			
			The visual and quantitative results of the experimental phantoms confirm that our proposed method improves the overall quality of the strain images.


			\subsubsection{Smoothing window effect on strain image}
\begin{figure}[!t]
				\centering
				\includegraphics[width=0.7\textwidth]{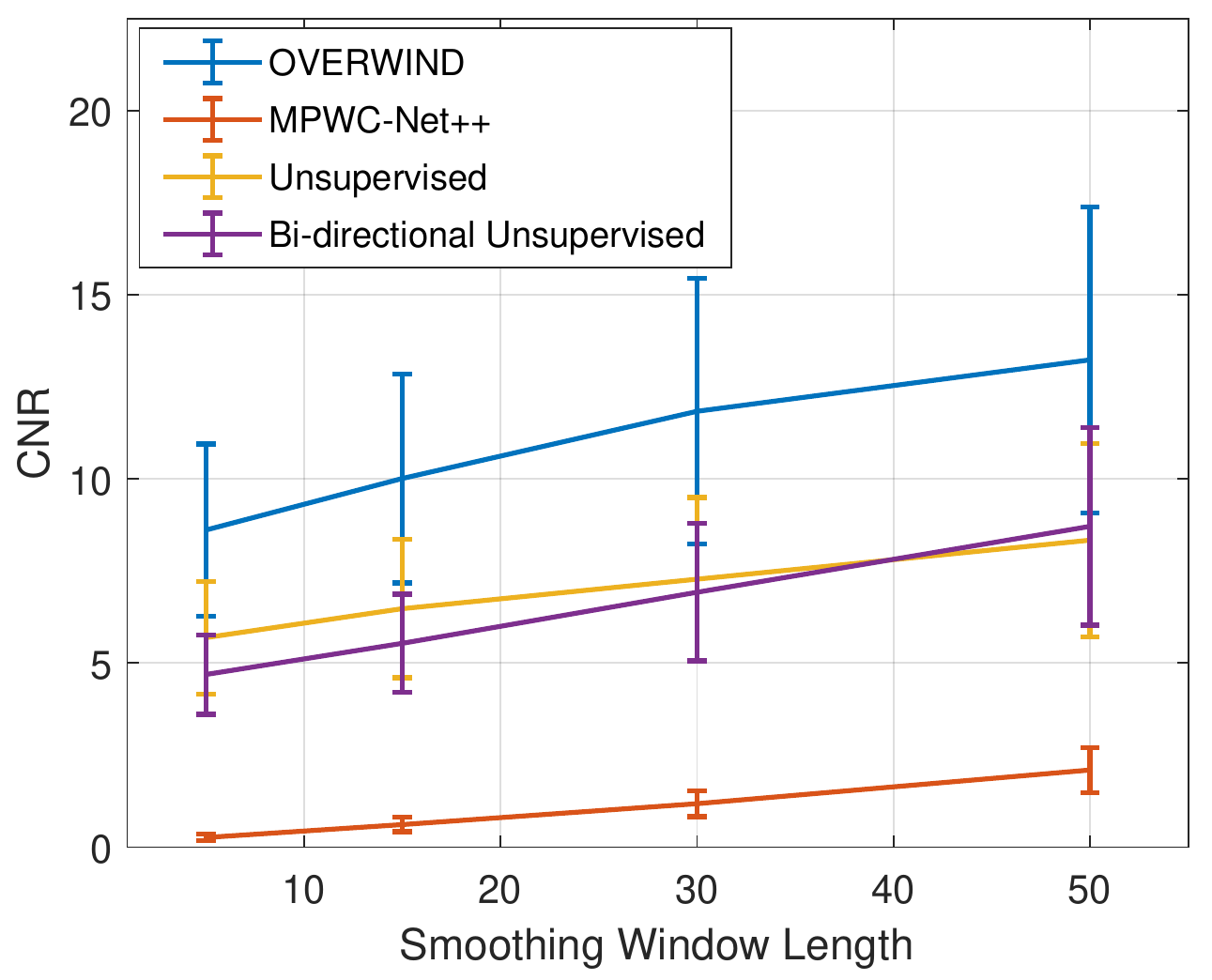}
				\centering
				\caption{CNR values of the compared method using different smoothing window lengths. The strain images are shown in Fig. \ref{fig:LSE}.}
				\label{fig:CNR_LSE}
			\end{figure}

\begin{figure}[!t]
				\centering
				\includegraphics[width=0.99\textwidth]{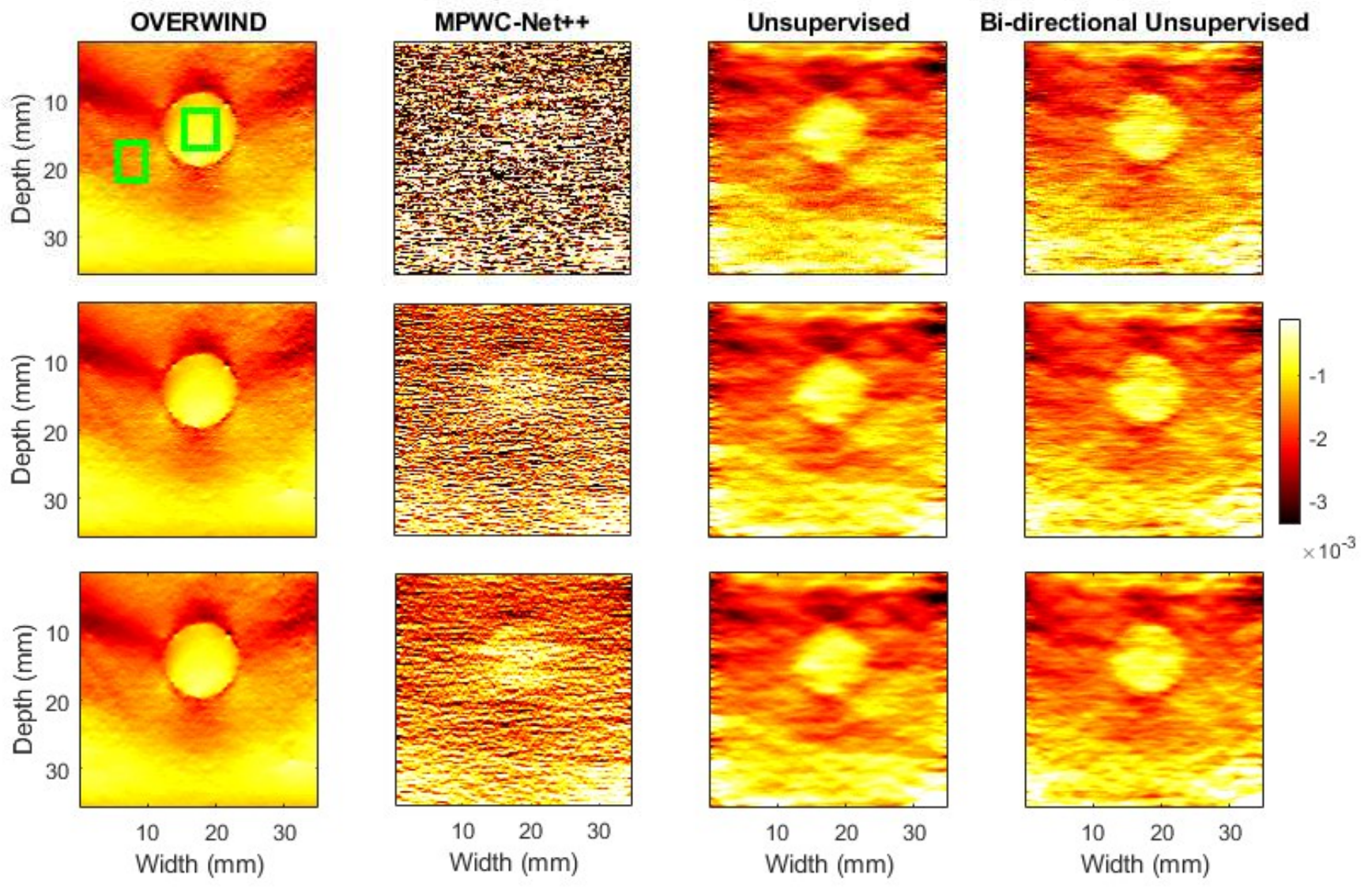}
				\centering
				\caption{The strain images of the compared methods. Smoothing window size is 5 (top), 15 (middle) and 30 (bottom).}
				\label{fig:LSE}
			\end{figure}

			After displacement estimation, a smoothing window along with the derivative kernel are used to reduce the error and compute the derivative of the displacement. Larger windows smooth the displacement more but sacrifice the resolution of the strain image. Therefore, displacement estimation methods that require smaller windows are preferred. When a USE displacement estimation method does not require a large smoothing window, it shows that the method produces a displacement map with a low variance error. To compare the methods, we compute the strain image of two image pairs with smoothing windows of sizes 5, 15, 30, and 40. The CNR values of different smoothing window lengths are shown in Fig. \ref{fig:CNR_LSE}. It can be seen that OVERWIND has high CNR values even when the smallest smoothing window is employed. Unsupervised and bi-directional unsupervised methods have good CNR values close to that of OVERWIND. It should be mentioned that the difference between the unsupervised and bi-directional methods in Fig. \ref{fig:CNR_LSE} was not statistically significant (\textit{p-}value $=0.112$). MPWC-Net++ has very low CNR when the smoothing window is small. It indicates that this method is highly sensitive to the length of smoothing window and requires larger ones to produce acceptable strain images, whereas OVERWIND and the two unsupervised methods do not need a large smoothing window to produce reliable strain images and have low variance errors. The strain images are shown in Fig. \ref{fig:LSE} for smoothing windows of 5 (top), 15 (middle) and 30 (bottom), and the target and background windows for computation of CNR are highlighted. We can see that MPWC-Net++ generates noisy strain images for small smoothing windows, where the inclusion is not visually detectable. However, both unsupervised fine-tuning methods provide a better performance, close to OVERWIND, and generate less variations compared to MPWC-Net++.
			
			\subsection{Lateral Strain}
\begin{figure}[!t]
				\centering
				\includegraphics[width=0.90\textwidth]{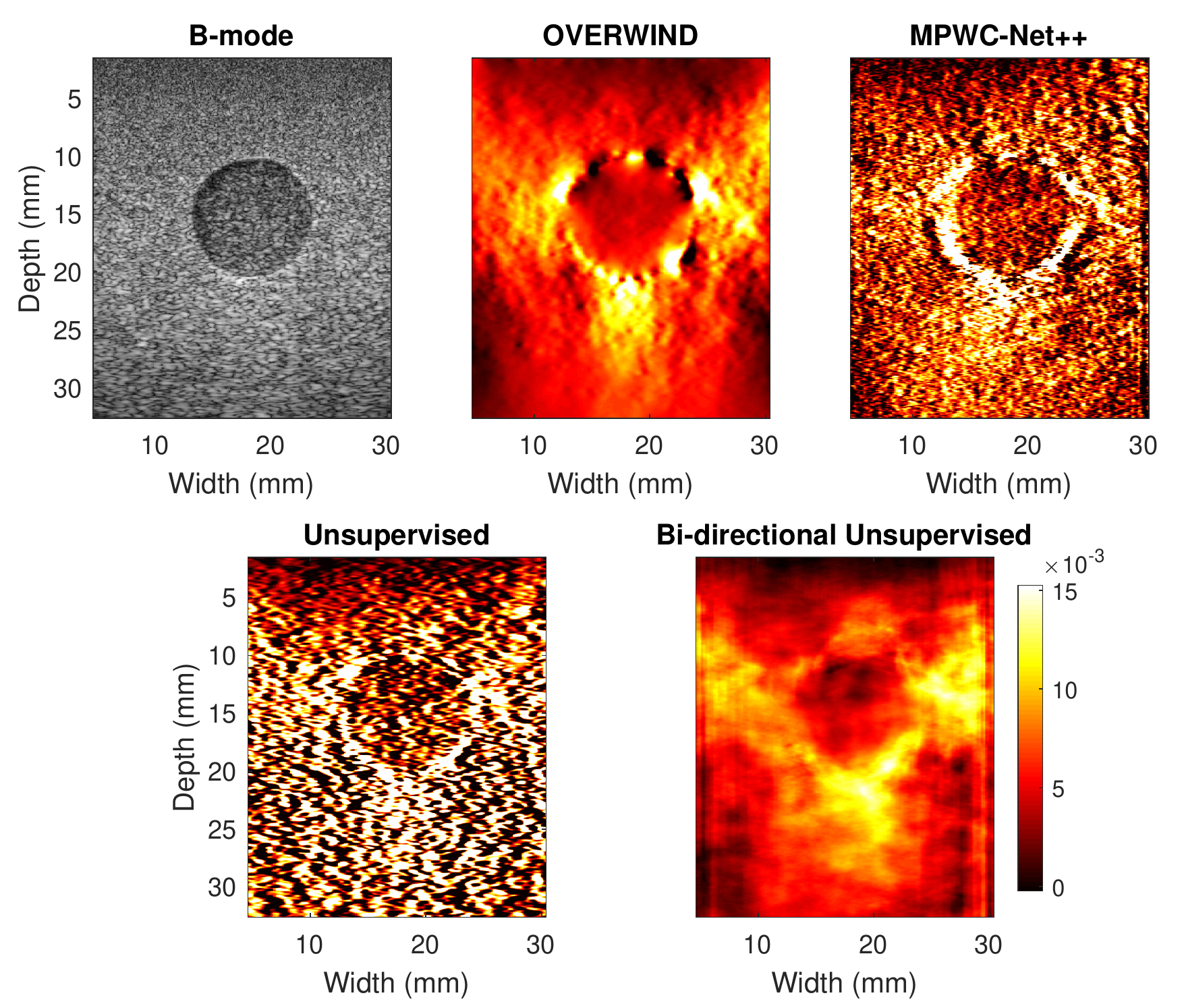}
				\centering
				\caption{The lateral strain results. The proposed bi-directional method provides a smooth strain image close to OVERWIND, while MPWC-Net++ and the unsupervised method generate noisy strain images and the inclusion is barely visible.}
				\label{fig:lateral}
			\end{figure}

			The lateral strain ($\frac{\partial W_l}{\partial l}$) has much lower quality than the axial strain since the main movement is in the axial direction, and the lateral sampling frequency and resolution are low. The lateral strain can be utilized in inverse problem methods to find the elastic modulus \cite{mohammadi2021ultrasound}. Fig. \ref{fig:lateral} shows the lateral strain obtained by the compared methods (refer to the Supplementary Materials for the axial strain). It can be seen that MPWC-Net++ and the unsupervised method obtain very noisy strain images and the inclusion is hardly visible. However the proposed bi-directional method and OVERWIND obtain acceptable strain images and the inclusion can be detected. 

			\begin{table*}[]
				\caption{Quantitative results of \textit{in vivo} data. The pairs marked by asterisk or star are not statistically significant (\textit{p}-value$>0.01$).} 
				\label{tab:invivo}
				\begin{tabular}{ccccc}
					\hline
					& \multicolumn{2}{c}{\textbf{\textit{In vivo} data 1}} & \multicolumn{2}{c}{\textbf{\textit{In vivo} data 2}} \\ \hline
					& CNR                  & SR(\%)              & CNR                 & SR(\%)                \\ 
					OVERWIND                    & \textbf{17.28$\pm$5.31}       & 21.80$\pm$4.16     & 7.18$\pm$1.58$^\ast$       & 51.40$\pm$6.06$^\star$        \\ \hline \addlinespace[0.125cm]
					MPWC-Net++                  & 11.51$\pm$3.08       & 25.50$\pm$6.63      & 7.31$\pm$2.47$^\ast$       & 48.89$\pm$11.61$^\ast$       \\ \addlinespace[0.125cm]
					Unsupervised                & 11.91$\pm$2.62       & 19.20$\pm$5.30      & 6.73$\pm$3.09      & 47.83$\pm$20.83$^\star$$^\ast$       \\\addlinespace[0.125cm]
					Bi-directional Unsupervised & \uline{16.27$\pm$5.26}       & 19.35$\pm$5.46      & 7.91$\pm$3.18       & \uline{\textbf{45.60$\pm$11.83}}       \\ \addlinespace[0.125cm]
					\multicolumn{1}{l}{Bi-directional Unsupervised + ft} &
					\multicolumn{1}{l}{14.37$\pm$4.40} &
					\multicolumn{1}{l}{\uline{\textbf{19.19$\pm$6.10}}} &
					\multicolumn{1}{l}{\uline{\textbf{8.86$\pm$2.64}}} &
					\multicolumn{1}{l}{46.14$\pm$12.91} \\ \hline
				\end{tabular}
			\end{table*}
			\subsection{\textit{In vivo} Results}
\begin{figure}[!t]
				\centering
				\includegraphics[width=0.95\textwidth]{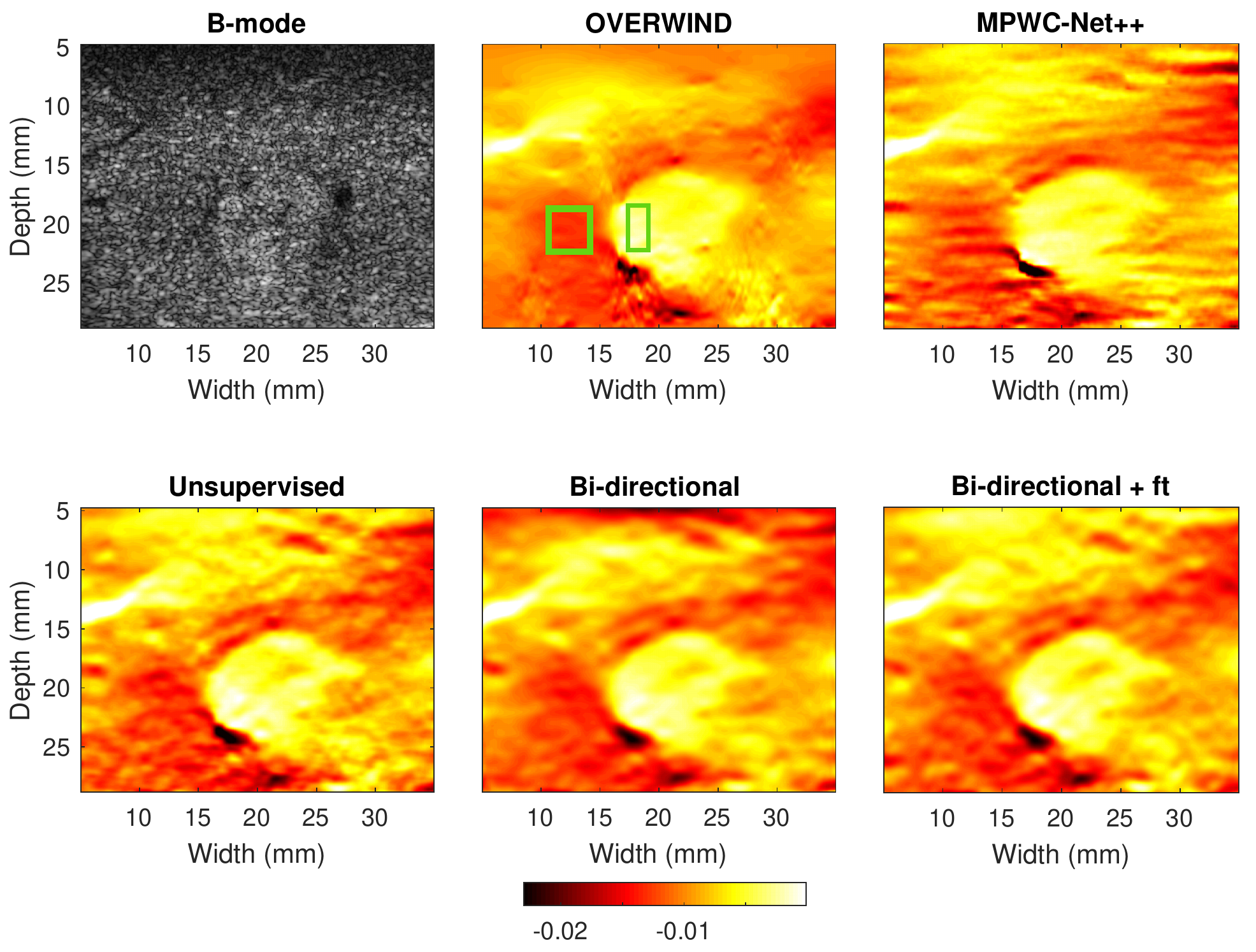}
				\centering
				\caption{Strain images of \textit{in vivo} data 1. The tumor has a lower absolute strain value but looks brighter since the strain is negative.}
				\label{fig:invivo1}
			\end{figure}

\begin{figure}[!t]
				\centering
				\includegraphics[width=0.95\textwidth]{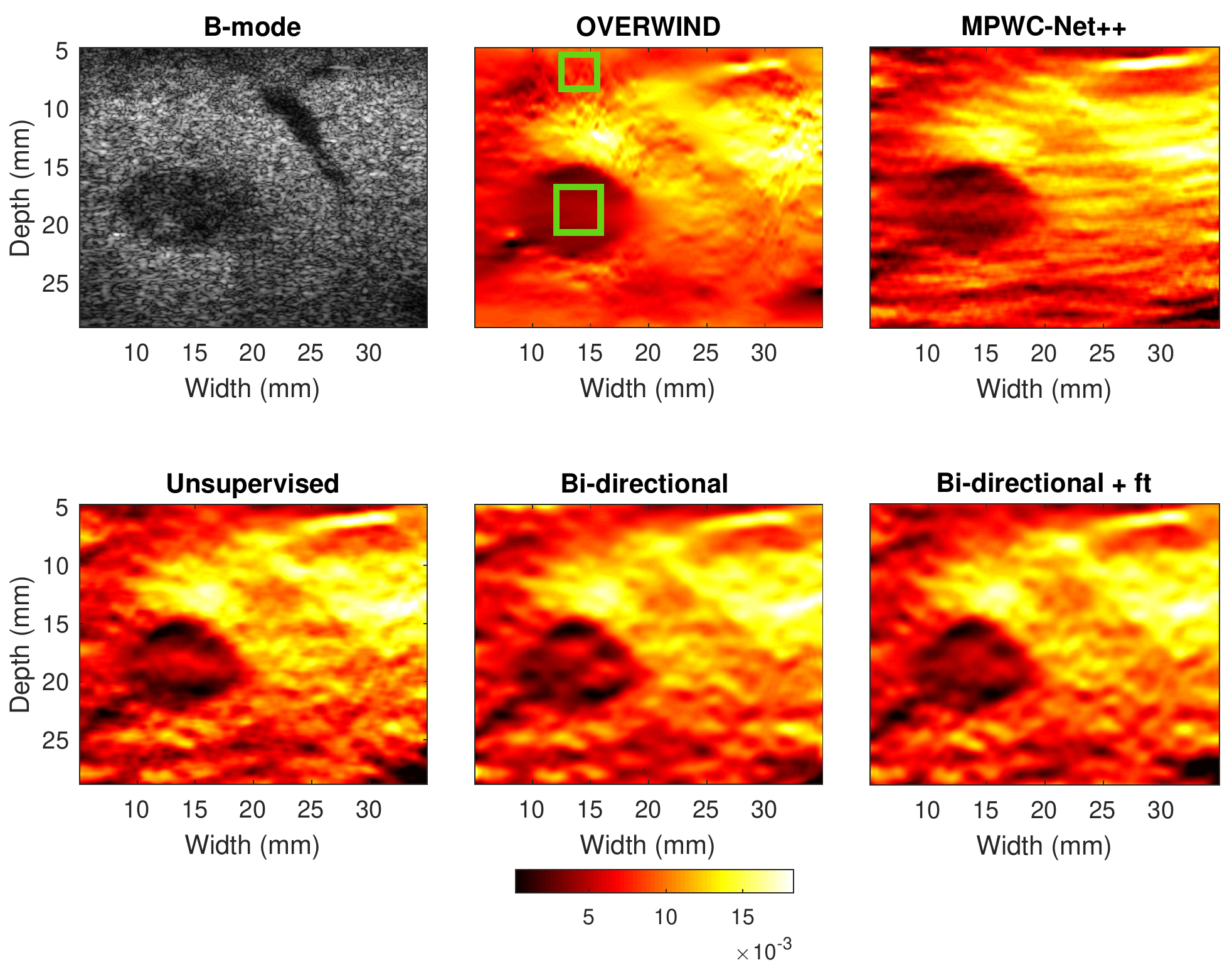}
				\centering
				\caption{Strain images of \textit{in vivo} data 2. The tumor has a lower absolute strain value and looks darker since the strain is positive.}
				\label{fig:invivo2}
			\end{figure}

			Compared methods are evaluated with two \textit{in vivo} data belonging to two patients. We also fine-tuned the bi-directional network using \textit{in vivo} data to find out if further improvements can be achieved. The strain images of the compared methods are given in Fig. \ref{fig:invivo1} and \ref{fig:invivo2}. OVERWIND produces high-quality strain images with low noise, while the strain images obtained by MPWC-Net++ have some over-smoothing especially in the lateral direction. Both unsupervised methods substantially improve the strain image qualities of MPWC-Net++. The bi-directional+ft also obtains high quality strain images, but the difference with the bi-directional method is not discernible.

			The quantitative results are given in Table \ref{tab:invivo}. OVERWIND achieves the highest CNR for \textit{in vivo} data 1, while bi-directional unsupervised+ft has the best CNR for \textit{in vivo} data 2. It can also be seen that fine-tuning on \textit{in vivo} data does not results in considerable CNR improvement (it has slightly better CNR than bi-directional for \textit{in vivo} data 2 and worse CNR for \textit{in vivo} data 1). In terms of SR, bi-directional and bi-directional+ft have the best SR values. MPWC-Net++ and OVERWIND have the highest SR among the compared methods.       
			
			\section{Discussion}
			In this paper, we employed semi-supervised training to improve the performance of an optical flow network for USE. Although we used MPWC-Net++, which outperformed other networks for USE, the training method can be applied to other networks as well. It should be mentioned that the optical flow networks usually have pyramidal structure meaning that the displacements are estimated in different resolutions. Similar to previous works \cite{tehrani2020displacement,tehrani2020semi} and unsupervised optical flow works \cite{jonschkowski2020matters}, we only used the last output resolution for fine-tuning.
			
			The optical flow CNNs trained on computer vision images do not employ regularizations due to abrupt changes in scenes such as a moving car in front of a fixed background. However, the displacement in USE is usually smooth without any sudden changes. The effect of the absence of the regularization can be seen in Fig. \ref{fig:LSE} (top), where the smoothing window is very small. The strain estimated by MPWC-Net++ has a high variance, whereas the bi-directional unsupervised method provides smooth and high-quality strain images by incorporating smoothness and consistency constrain in the estimated displacement.  
			
			The values reported in Tables \ref{tab:phantom_CNR}, \ref{tab:phantom_SR} and \ref{tab:invivo} are the mean and standard deviation of CNR and SR values. We have conducted statistical analysis to find out that if the difference between obtained values is statistically significant. We employ Friedman test \cite{theodorsson1987friedman}, and the \textit{p}-values are given in the Supplementary Materials.
			
			Regarding the choice of weights, the explained method of tuning the weights only gives a rough estimate of the optimum weights. However, the output strain image is not considerably sensitive to these weights and similar performance can be obtained by different weights.

			\begin{table}[]
				\caption{SSIM ($\%$) of strain images of OVERWIND vs deep learning methods.}
				\label{tab:ssim}
				\resizebox{0.99\textwidth}{!}{
					\begin{tabular}{@{}lllllll@{}}
						\toprule
						& Fig. \ref{fig:phantom1} & Fig. \ref{fig:phantom2} & Fig. \ref{fig:phantom3} & Fig. \ref{fig:phantom4}                     & \textit{in vivo} 1 & \textit{in vivo} 2 \\ \midrule
						MPWC-Net++                  & 90.12  & 73.36  & 60.84  & \multicolumn{1}{l|}{90.77} & 92.66    & 93.79     \\
						Unsupervised                & \textbf{98.96}  & 95.56  & 94.86  & \multicolumn{1}{l|}{96.11} & 90.08    & 91.13     \\
						Bi-directional Unsupervised & 98.89  & \textbf{96.27}  & \textbf{95.76}  & \multicolumn{1}{l|}{\textbf{96.90}} & \textbf{93.57}    & \textbf{96.62}     \\ \bottomrule
				\end{tabular}}
			\end{table}
			To consider texture (which is not captured by SR or CNR), structural similarity index (SSIM) \cite{Brunet2011} has been used to evaluate the texture similarity between OVERWIND and the deep learning methods. The results are provided in Table \ref{tab:ssim}. It is clear that the proposed bi-directional method achieves a higher SSIM score compared to the other methods which indicates that it produces strain images closer to OVERWIND than the other compared methods.
			
			Running time is another important aspect that needs to be investigated. The deep learning methods shine in this aspect, our proposed method can provide high quality strain images close to OVERWIND for real-time applications. To give a general view about the computation time, OVERWIND takes 26 seconds for an image pair of size $1920\times384$ on CPU ($8^{th}$ generation, core i7). It should be noted that this is a Matlab implementation, and an optimized implementation in C will be much faster. MPWC-Net++ and bi-directional MPWC-Net++ take 0.166 and 0.174 seconds on NVIDIA A100 GPU, respectively. The bi-directional variant of MPWC-Net++ takes slightly more time than MPWC-Net++ since it estimates both forward and backward displacements; however, it still takes much less time than running the network two times and computing forward and backward displacements separately.

			\section{Conclusion}
			In this paper, we proposed a bi-directional semi-supervised deep learning method. We used strain consistency along with data and smoothness loss. We also employed the second-order derivatives regularization of axial and lateral displacements in both directions. Furthermore, we reduced the number of hyper-parameters by relating some of them to others by taking into account the underlying principles of the US. We showed that our proposed method substantially improved current optical flow networks used for USE. We validated our method using different experimental phantom and \textit{in vivo} data. Our proposed method obtained strain images close to OVERWIND.

			\section*{Acknowledgment}
			
			We thank the anonymous reviewers and the handling
			Associate Editor for constructive comments. We acknowledge the support of the Natural Sciences
			and Engineering Research Council of Canada (NSERC)
			RGPIN-2020-04612.


			
			
			%
			\bibliography{IEEEfull}
			\bibliographystyle{IEEEtran}
			
		\end{document}


\title{Supplementary Material for Bi-Directional Semi-Supervised Training of \\Convolutional Neural Networks for \\Ultrasound Elastography Displacement Estimation}
	\author{Ali~K. Z. Tehrani, Mostafa~Sharifzadeh, Emad Boctor and Hassan~Rivaz 
		
		\thanks{A. K. Z. Tehrani, Mostafa Sharifzadeh, and H. Rivaz are with the Department
			of Electrical and Computer Engineering, Concordia University, Canada,
			Emad Boctor is with the Department of Radiology and Radiological Science, John Hopkins University, United States,
			e-mail:A\_Kafaei@encs.concordia.ca,
			mostafa.sharifzadeh@concordia.ca,
			eboctor@jhmi.edu and 
			hrivaz@ece.concordia.ca
		}
		\thanks{}}

\maketitle


	\section{Statistical Analysis}
	In Table. I, II and III, the mean and standard deviation values of CNR and SR of the small patches inside the windows highlighted in the corresponding figures are reported. Here, we investigate whether the obtained values represent statistically significant differences. The CNR and SR of small patches are not normally distributed; therefore, we utilize the Friedman test. The \textit{p}-values of this test are reported for CNR and SR in Table \ref{tab:cnr_pvalue} and \ref{tab:sr_pvalue} in Supplementary Materials, respectively. The bold values indicate \textit{p}-values smaller than 0.01. It can be observed that a large majority of pairs are significantly different. The mean value of CNR of OVERWIND and bi-directional method of Fig. 2 reported in Table I are very close (27.26 and 27.71) but now it can be seen that they are not significantly different. The same is true for other results, which are not significantly different.   

	\begin{table*}[]
		\renewcommand\thetable{SI}
		\caption{The \textit{p}-values of Friedman test of significant difference of CNR values. The bold font denotes statistical significance (\textit{p}-value$<0.01$) is achieved.}
		\label{tab:cnr_pvalue}
		
		\resizebox{0.95\textwidth}{!}{%
			\begin{tabular}{@{}cccccccc@{}}
				\toprule
				&
				Fig. 2 &
				Fig. 3 &
				Fig. 4 &
				Fig. 5(1) &
				Fig. 5(2) &
				\textit{in vivo} 1 &
				\textit{in vivo} 2 \\ \midrule
				{\color[HTML]{000000} OVERWIND vs MPWC-Net++} &
				{\color[HTML]{000000} \textbf{4.9e-139}} &
				{\color[HTML]{000000} \textbf{0}} &
				{\color[HTML]{000000} \textbf{1.1e-184}} &
				{\color[HTML]{000000} \textbf{0}} &
				{\color[HTML]{000000} \textbf{0}} &
				{\color[HTML]{000000} \textbf{7.9e-198}} &
				{\color[HTML]{000000} 0.084} \\
				\rowcolor[HTML]{CBCEFB} 
				{\color[HTML]{000000} OVERWIND vs Unsupervised} &
				{\color[HTML]{000000} \textbf{5.3e-22}} &
				{\color[HTML]{000000} \textbf{3.3e-283}} &
				{\color[HTML]{000000} \textbf{6.6e-71}} &
				{\color[HTML]{000000} \textbf{0}} &
				{\color[HTML]{000000} \textbf{0}} &
				{\color[HTML]{000000} \textbf{1.2e-201}} &
				{\color[HTML]{000000} \textbf{1.7e-7}} \\
				{\color[HTML]{000000} OVERWIND vs Bi-directional Unsupervised} &
				{\color[HTML]{000000} 0.47} &
				{\color[HTML]{000000} \textbf{6.4e-130}} &
				{\color[HTML]{000000} \textbf{1.0e-8}} &
				{\color[HTML]{000000} 0.123} &
				{\color[HTML]{000000} \textbf{5.7e-171}} &
				{\color[HTML]{000000} \textbf{2.6e-11}} &
				{\color[HTML]{000000} \textbf{2.8e-53}} \\
				\rowcolor[HTML]{CBCEFB} 
				{\color[HTML]{000000} Unsupervised vs MPWC-Net++} &
				{\color[HTML]{000000} \textbf{4.9e-139}} &
				{\color[HTML]{000000} \textbf{3.4e-320}} &
				{\color[HTML]{000000} \textbf{1.6e-179}} &
				{\color[HTML]{000000} \textbf{1.8e-199}} &
				{\color[HTML]{000000} \textbf{0}} &
				{\color[HTML]{000000} \textbf{0.003}} &
				{\color[HTML]{000000} \textbf{1.9e-12}} \\
				{\color[HTML]{000000} Bi-directional Unsupervised vs MPWC-Net++} &
				{\color[HTML]{000000} \textbf{4.9e-139}} &
				{\color[HTML]{000000} \textbf{0}} &
				{\color[HTML]{000000} \textbf{1.1e-184}} &
				{\color[HTML]{000000} \textbf{0}} &
				{\color[HTML]{000000} \textbf{0}} &
				{\color[HTML]{000000} \textbf{1.38e-167}} &
				{\color[HTML]{000000} \textbf{9.6e-39}} \\
				\rowcolor[HTML]{CBCEFB} 
				{\color[HTML]{000000} Bi-directional Unsupervised vs Unsupervised} &
				{\color[HTML]{000000} \textbf{8.7e-37}} &
				{\color[HTML]{000000} \textbf{4.5e-39}} &
				{\color[HTML]{000000} \textbf{1.3e-61}} &
				{\color[HTML]{000000} \textbf{5.1e-269}} &
				{\color[HTML]{000000} \textbf{0}} &
				{\color[HTML]{000000} \textbf{5.4e-159}} &
				{\color[HTML]{000000} \textbf{2.7e-196}} \\ \bottomrule
		\end{tabular}}
	\end{table*}

	\begin{table*}[]
		\renewcommand\thetable{SII}
		\caption{The \textit{p}-values of Friedman test of significant difference of SR values. The bold font denotes statistical significance (\textit{p}-value$<0.01$) is achieved. }
		\label{tab:sr_pvalue}
		
		\resizebox{0.95\textwidth}{!}{%
			\begin{tabular}{@{}cccccccc@{}}
				\toprule
				&
				Fig. 2 &
				Fig. 3 &
				Fig. 4 &
				Fig. 5(1) &
				Fig. 5(2) &
				\textit{in vivo} 1 &
				\textit{in vivo} 2 \\ \midrule
				{\color[HTML]{000000} OVERWIND vs MPWC-Net++} &
				{\color[HTML]{000000} \textbf{5.3e-21}} &
				{\color[HTML]{000000} \textbf{6.8e-152}} &
				{\color[HTML]{000000} \textbf{4.9e-8}} &
				{\color[HTML]{000000} \textbf{3.38e-153}} &
				{\color[HTML]{000000} \textbf{5.03e-17}} &
				{\color[HTML]{000000} \textbf{4.5e-93}} &
				{\color[HTML]{000000} \textbf{4.6e-15}} \\
				\rowcolor[HTML]{CBCEFB} 
				{\color[HTML]{000000} OVERWIND vs Unsupervised} &
				{\color[HTML]{000000} \textbf{5.5e-89}} &
				{\color[HTML]{000000} \textbf{0}} &
				{\color[HTML]{000000} 0.0324} &
				{\color[HTML]{000000} 0.076} &
				{\color[HTML]{000000} \textbf{3.3e-107}} &
				{\color[HTML]{000000} \textbf{4.7e-16}} &
				{\color[HTML]{000000} 0.215} \\
				{\color[HTML]{000000} OVERWIND vs Bi-directional Unsupervised} &
				{\color[HTML]{000000} \textbf{4.9e-139}} &
				{\color[HTML]{000000} \textbf{0}} &
				{\color[HTML]{000000} \textbf{1.6e-85}} &
				{\color[HTML]{000000} \textbf{1.1e-241}} &
				{\color[HTML]{000000} \textbf{0}} &
				{\color[HTML]{000000} \textbf{3.4e-51}} &
				{\color[HTML]{000000} \textbf{1.3e-111}} \\
				\rowcolor[HTML]{CBCEFB} 
				{\color[HTML]{000000} Unsupervised vs MPWC-Net++} &
				{\color[HTML]{000000} 0.038} &
				{\color[HTML]{000000} \textbf{1.8e-223}} &
				{\color[HTML]{000000} 0.836} &
				{\color[HTML]{000000} \textbf{2.7e-58}} &
				{\color[HTML]{000000} \textbf{1.5e-99}} &
				{\color[HTML]{000000} \textbf{4.9e-96}} &
				{\color[HTML]{000000} 0.0335} \\
				{\color[HTML]{000000} Bi-directional Unsupervised vs MPWC-Net++} &
				{\color[HTML]{000000} \textbf{6.4e-87}} &
				{\color[HTML]{000000} \textbf{0}} &
				{\color[HTML]{000000} \textbf{1.4e-33}} &
				{\color[HTML]{000000} \textbf{0}} &
				{\color[HTML]{000000} \textbf{1.6e-286}} &
				{\color[HTML]{000000} \textbf{1.2e-201}} &
				{\color[HTML]{000000} \textbf{2.2e-23}} \\
				\rowcolor[HTML]{CBCEFB} 
				{\color[HTML]{000000} Bi-directional Unsupervised vs Unsupervised} &
				{\color[HTML]{000000} \textbf{4.9e-139}} &
				{\color[HTML]{000000} \textbf{5.2e-35}} &
				{\color[HTML]{000000} \textbf{1.3e-11}} &
				{\color[HTML]{000000} \textbf{4.2e-35}} &
				{\color[HTML]{000000} \textbf{6.0e-7}} &
				{\color[HTML]{000000} \textbf{9.0e-17}} &
				{\color[HTML]{000000} \textbf{1.9e-12}} \\ \bottomrule
		\end{tabular}}
	\end{table*}

	\subsection{Axial Strain and Lateral Displacement of Fig. 8}
\begin{figure*}[!t]
		\renewcommand\thefigure{S1}
		\centering
		\includegraphics[width=0.55\textwidth]{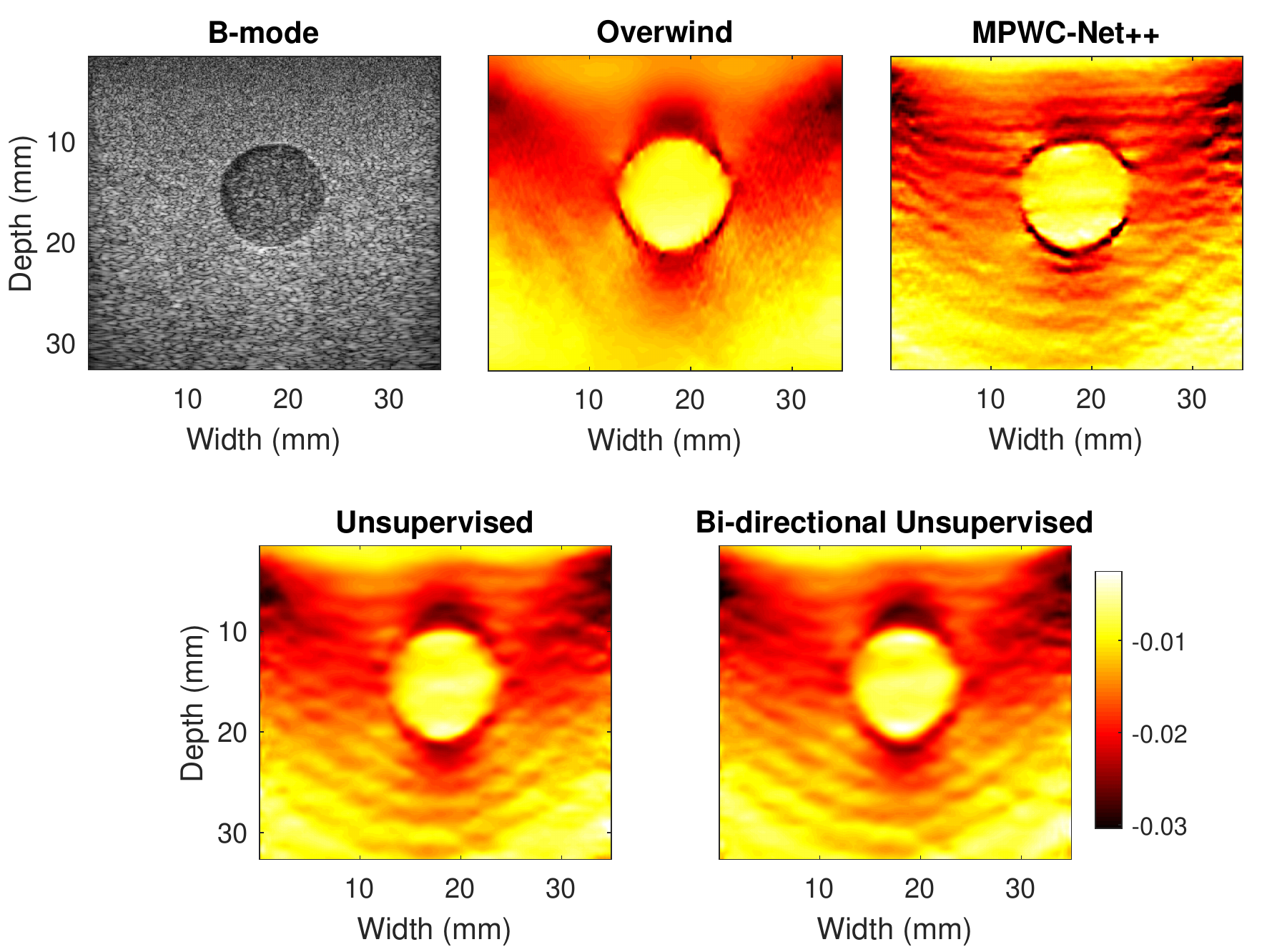}
		\centering
		\caption{The axial strain results. }
		\label{fig:axial}
	\end{figure*}

	The lateral strain of the experimental phantom image pair is shown in the paper. Here the axial strain results are shown in Fig. \ref{fig:axial}. The axial strain provides higher quality strain images.
